\documentclass[12pt]{book}

\usepackage{makeidx,tsukuba}
\usepackage{graphicx}

\makeauthorindex
\makeindex

\begin{document}

\BookTitle{}
\CopyRight{}
\pagenumbering{arabic}

\chapter{
Particle Acceleration in Supernova Remnants in the\\ Presence of Streaming
Instability and Nonlinear Wave\\ Interactions}

\author{
%
%
V.S. Ptuskin and V.N. Zirakashvili\\
{\it IZMIRAN, Russian Academy of Sciences, Troitsk, Moscow region 142190, 
Russia}}

\section*{Abstract}

The instability in the cosmic-ray (CR) precursor of a SN shock is studied. 
The
level of turbulence in this region determines the maximum energy of 
accelerated CRs.
The consideration is not limited by the case of weak
turbulence. It is assumed that Kolmogorov type nonlinear wave interactions
together with the ion-neutral collisions restrict the amplitude of random
magnetic field. As a result, the maximum energy of accelerated particles
strongly depends on the age of a SNR. It can be as high as $10^{17}Z$ eV in
young SNR and falls down to about $10^{10}Z$ eV at the end of Sedov stage 
($Z$
is the particle charge). This finding may explain why the SNRs with the age
more than a few thousand years are not prominent sources of very high energy
$\gamma$-rays. The averaged spectrum of ultrarelativistic CR injected in the
interstellar medium is close or somewhat steeper than $E^{-2}$.

\section{Introduction}

The dependent on energy diffusion coefficient $D(E)$ determines the maximum
energy that particles can gain in the process of acceleration. The condition
of efficient acceleration is $D(E)\leq ku_{\mathrm{sh}}R_{\mathrm{sh}}$,
where $R_{\mathrm{sh}}$ is the radius and $u_{\mathrm{sh}}$ is the velocity 
of
spherical shock, $k=0.1$ in the free expansion stage and $k=0.04$ in Sedov
stage, e.g. [3]. The Bohm value $D_{B}=vr_{\mathrm{g}}/3$
($v$ is the particle velocity, and $r_{g}$ is the particle Larmor radius) 
that
is a lower bound of the diffusion along the magnetic field gives $E_{\max
}=1.7\times10^{14}Z\left(  \mathcal{E}_{51}/n_{0}\right)  $ eV at the end of
the free expansion stage when particles reach the highest energy. Here the 
SN
burst with kinetic energy of ejecta $\mathcal{E=E}_{51}10^{51}$ erg in the 
gas
with density $n_{0}$ cm$^{{-3}}$ and the interstellar magnetic field
$B_{0}=5\times10^{-6}$ G is considered.

Analyzing the early stage of SNR evolution when the shock velocity is high,
$u_{\mathrm{sh}}\sim10^{4}$ km s$^{-1}$, it was found [2] that
the CR streaming instability can be so strong that the amplified field 
$\delta
B\geq10^{-4}$ G far exceeds $B_{0}$. The maximum particle energy increases
accordingly. The CR streaming instability is less efficient as the shock
velocity decreases with time and the nonlinear wave interactions reduce the
level of turbulence at the late Sedov stage [7,10].
This leads to fast diffusion and decreases $E_{\max}$. The effect is
aggravated by the possible wave damping on the ion-neutral collisions [1,6].
In the present work, we consider the acceleration of
CR and their streaming instability in a wide range of shock velocities. The
level of magnetic field fluctuations is allowed to be arbitrarily large,
and the rate of nonlinear wave interactions is
assumed to correspond to the Kolmogorov nonlinearity. The collisional
dissipation is also taken into account. The task is to find the
maximum energy of accelerated particles as a function of SNR age.

\section{Maximum Energy of Accelerated Particles}

In the test particle approximation, the distribution of particles in 
momentum
for high Mach number shocks has the canonical form $f(p)\sim p^{-4}$. In
the case of efficient acceleration, the action of CR pressure on the shock
structure causes nonlinear modification of the shock that changes the shape 
of
particle spectrum making it flatter at relativistic energies. So, we assume
that the distribution at the shock is of the form $f_{0}(p)\sim
p^{-4+a}$ where $0<a<0.5$, and value $a=0.3$ is used in the numerical
estimates below. The normalization of function $f_{0}(p)$ is such that the
integral $N=4\pi\int dpp^{2}f_{0}(p)$ gives the number density of CR. We
assume that the CR pressure at the shock is some fraction $\xi_{\mathrm{cr}%
}\leq1$ of the upstream momentum flux entering the shock front, so that
$P_{\mathrm{cr}}=\xi_{\mathrm{cr}}\rho u_{\mathrm{sh}}^{2}$. The typical 
value
of $\xi_{\mathrm{cr}}=0.5$ and the total compression ratio $7$ were found 
for
strongly modified shocks in [3].

The following steady-state equation determines the energy density $W$ of the
turbulence amplified by the streaming instability in the CR precursor 
upstream
of the shock:
\begin{equation}
u\nabla W=2(\Gamma_{\mathrm{cr}}-\Gamma_{\mathrm{l}}-\Gamma_{\mathrm{nl}%
})W.%
\end{equation}
Here the l.h.s. describes the advection of turbulence by supersonic gas 
flow.
The terms on the r.h.s. of the equation describe respectively the wave
amplification by CR, the linear damping of waves in background plasma and 
the
nonlinear wave-wave interactions that limit the amplitude of turbulence. The
Kolmogorov-type nonlinearity with the simplified expression $\Gamma
_{\mathrm{nl}}=(2C_{\mathrm{K}})^{-3/2}V_{\mathrm{a}}kA(>k)\approx
0.05V_{\mathrm{a}}kA(>k)$ at $C_{\mathrm{K}}=3.6$ (as given by the numerical
simulations [9]) is used in our calculations. Here $A=\delta
B/B_{0}$ is the wave amplitude, $k$ is the wave number. The wave-particle
interaction is of resonant character and the resonance condition is
$kr_{\mathrm{g}}=1$. The equations for $\Gamma_{\mathrm{cr}}$ and for $D$ 
used
in our calculations generalize the standard equations [4]
derived in the case of weak random field. The details of our
consideration can be found in [8].

\begin{figure}[t]
\begin{center}
\includegraphics[height=20pc]{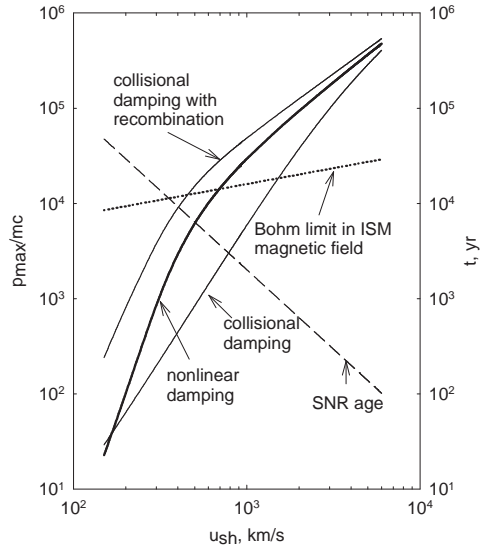}
\end{center}
\par
\vspace{-0.5pc}
\caption{The particle maximum momentum $p_{\max}$ in units $mc$
as a function of shock velocity $u_{\mathrm{sh}}$.}
\end{figure}

Fig.1 illustrates the results of calculations of $E_{\max}$ at the Sedov 
stage
of SNR evolution at $\mathcal{E}=10^{51}$ erg in the warm interstellar gas
with the temperature $T=8\times10^{3}$ K, and the average density 
$n_{0}=0.4$
cm$^{-3}$. Three solid lines correspond to three cases of wave dissipation
considered separately: nonlinear wave interactions; damping by ion-neutral
collisions at constant gas density; damping by ion-neutral collisions when 
the
diffuse neutral gas restores its density after complete ionization by the
radiation from SN burst. The maximum energy of protons accelerated by SN
shocks at the early Sedov stage is close to $3\times10^{14}$ eV (that 
exceeds
the Bohm limit) and decreases to about $10^{10}$ eV at the end of the Sedov
stage (that is less than the Bohm limit). In particular, the particle energy
is less than $10^{13}$ eV for $t>3\times10^{3}$ yr and this may explain the
absence of a TeV $\gamma$-ray signal from many SNRs [5]
where $\gamma$-rays could be produced through $\pi^{0}$ decays if 
sufficiently
energetic CRs were present.

The highest particle energy estimated as $E_{\max}=2\times10^{17}%
Z(u_{\mathrm{sh}}/3\times10^{4}$ km s$^{-1})^{2}\xi_{\mathrm{cr}%
}M_{\mathrm{ej}}^{1/3}n^{1/6}$ eV is reached at the end of the free 
expansion
stage ($M_{\mathrm{ej}}$ is the mass of ejecta in solar masses).

\section{Average Spectrum of Injected Cosmic Rays}

Let us find the overall spectrum of CR injected to the interstellar space. 
It
is not clear yet how the process of CR exit from the SNRs proceeds.
Consider two cases:

{\it A}) All accelerated particles leaves the envelope without considerable
adiabatic losses. The production of CR in the galactic disk is then
$Q=\nu_{\mathrm{sn}}4\pi\int dtu_{_{\mathrm{sh}}}R_{\mathrm{sh}}^{2}f_{0}$,
where $\nu_{\mathrm{sn}}$ is the SN rate in the Galaxy. This gives the
approximate scaling $Q\sim \nu_{\mathrm{sn}}\xi_{\mathrm{cr}}
\mathcal{E}p^{-4}$ (at $p\gg mc$) under the conditions
that $\xi_{\mathrm{cr}}=$ const, and that SNR expansion is adiabatic as in 
the case of
Sedov solution. It is remarkable that the form of average source spectrum 
$Q(p)$
is not sensitive to parameter $a$ at $a>0$ (see also simulations [3]).

{\it B}) The CR with maximum energies can not be confined near the shock
and are leaving the precursor because of the $p_{\max}(t)$ \ decrease. Other
particles which are confined in the envelope experience adiabatic energy
changes until the very end of Sedov stage and the exit from a SNR. The 
runaway
particles has the average spectrum of the $\ $form $p^{-4}$, whereas the
trapped particles gives approximately $p^{-4.1}$. The forms of these  
spectra are not
sensitive to $a$ at $a>0$, and both populations have the comparable number 
densities.

\section{Conclusion}

The accounting for non-linear effects in the instability that accompanies 
the
CR acceleration may simultaneously eliminate two difficulties of modern 
cosmic
ray astrophysics. It raises the maximum energy of accelerated particles in
young SNR above the standard Bohm limit through the production of strong
random magnetic fields and thus helps to explain the origin of galactic CR
with energies up to $10^{17}Z$ eV. It also decreases the maximum energy of
particles in the late Sedov stage of SNR evolution, which allows us to 
explain
why these objects are not bright in very high energy $\gamma$-rays. It is
remarkable that even with a flat instantaneous particle spectrum at the 
shock
typical for the strongly nonlinear regime of acceleration, the spectrum of
ultrarelativistic particles injected into interstellar space and averaged 
over
the age of SNR is close to $E^{-2}$ or somewhat steeper.

\textbf{Acknowledgments}. This work was supported by the RFBR grant.

\section{References}

%
%
%
%
%
%
%

\re
1.\ Bell A.R. 1978, MNRAS 182, 147\re
2.\ Bell A.R., Lucek S.G. 2001, MNRAS 321, 433\re
3.\ Berezhko E.G., Yelshin V.K., Ksenofontov L.T. 1996, JETP 82, 1\re
4.\ Berezinskii V.S. et al. 1990, Astrophysics of Cosmic Rays, (Amsterdam:
North Holland)\re
5.\ Buckley J.H. et al. 1998, A\&A 329, 639\re
6.\ Drury L. O'C., Duffy P., Kirk J.G. 1996, ApJ 309, 1002\re
7.\ Fedorenko V.N. 1990, Preprint 1442, A.F.Ioffe Phys. Tech. Inst., 
Leningrad\re
8.\ Ptuskin V.S., Zirakashvili V.N. 2003, astro-ph/0302053\re
9.\ Verma M.K. et al. 1996, JGR 101, 21619\re
10.\ V\"{o}lk H.J., Zank L.A., Zank G.P. 1988, A\&A 198, 274

\endofpaper
\end{document}